# Observation of Disorder State Coupling to Excitons in InGaN Disks in GaN Nanowires

Cameron Nelson *£†§, Yong-Ho Ra ‡, Zetian Mi £, Duncan Steel £†

£Department of Electrical Engineering and Computer Science, The University of Michigan, 1301 Beal Ave, Ann Arbor, MI 48109

†The H.M. Randall Laboratory of Physics, The University of Michigan, 450 Church Street, Ann Arbor, MI 48109

‡Department of Electrical and Computer Engineering, McGill University 3480 University Street, Montreal, Quebec H3A 0E9, Canada

**ABSTRACT:** $In_xGa_{1-x}N$ disks in GaN nanowires (DINWs) have emerged as a viable technology for on-chip tunable visible spectrum emission without the use of a phosphor. Here we present a study of the optical emission and absorption dynamics in DINWs that incorporates the important role of background disorder states. We show that the optical emission in the system is dominated by quantum-confined excitons, however the excitons are coupled to a large density of background disorder states. Rapid non-radiative decay (compared to other decay rates such as spontaneous emission) from disorder states into excitons is observed after optical excitation of our sample, which can be advantageous for increasing the brightness of the system in future design efforts.

The development of solid state light emitters without the use of a phosphor has been the subject of intense research interest over the last few decades. In principle, epitaxially grown layers of $In_xGa_{1-x}N$/GaN can be engineered to span the full visible emission spectrum by varying the InN concentration x. In conventional planar quantum well structures the growth of layers with full visible spectrum emission has been challenging due to issues such as efficiency droop at high injection currents *(1)*, large internal polarization fields *(2-3)* and rapidly decreasing emission efficiency as a function of increasing InN concentration, known as the "green gap" *(4-6)*. Many of these issues have been mitigated in InGaN disk-in-nanowire (DINW) structures that can be grown defect-free on foreign substrates such as Si *(7)* with significantly reduced internal electric fields due to strain relaxation in the active area *(8-9)*. These improvements have paved the way for recent demonstrations such as on-chip full visible spectrum lasers *(10-11)* and red-green-blue LED arrays *(7, 12-14)*.

The nature of optical emission from InGaN/GaN DINWs with diameters ~200 nm or less has been shown to be fundamentally different from InGaN/GaN quantum wells in planar heterostructures *(2,15)*. Smaller diameter DINWs support emission from exciton states (Coulomb-bound electron-hole pairs) that are confined to the center of the InGaN disk due to strain relaxation at the DINW side walls *(2,9,15-16)*. For planar InGaN quantum well light emitters, excitonic emission is typically quenched due to a large density of threading defects in the active region that are not present in DINWs *(8,17)*. Emission in planar quantum wells is instead provided by radiative recombination of localized electron-hole-pair energy states formed by disorder that are small enough to avoid the defect sites *(17)*. Disorder states arise naturally in InGaN systems from random variations in the InN concentration within the material or from atomistic defects *(17-19)* such as In-N-In chains *(19)*. The quantum efficiency of DINWs that support excitonic emission has been shown to considerably exceed that of planar quantum wells that are grown under similar conditions *(2)*, indicating that the DINW excitons are generally brighter compared to the disorder states. Therefore, in addition to full visible spectrum emission tunability, LED devices made from DINWs have shown the potential for large increases in emission efficiency over more conventional planar InGaN quantum well devices at all wavelengths *(2,7,14-16)*.

Recent results have confirmed that large densities of localized disorder states can also exist in DINW systems *(20)*, although, as noted above, the emission from DINWs has been shown to be dominated by radiative recombination from quantum confined exciton states. While there has been a large focus on characterization of excitons in DINWs, very little is known about the role of localized disorder states in the system.

In this work we study the linear and nonlinear optical properties of an ensemble of red emitting, selective area InGaN/GaN DINWs. First, using the intensity-dependent photoluminescence (PL) spectrum, we find that the sample emission is dominated by exciton states, however the linear absorption spectrum measured using photoluminescence excitation (PLE) shows that a large density of background disorder states are also present in the system. Inhomogeneous broadening from the background disorder states gives a nearly featureless, exponential PLE spectrum as a function of excitation energy.

Furthermore, it has been shown in previous reports that the nonlinear absorption ($\chi^{(3)}$) spectrum can provide a sensitive filter for probing the optical physics of exciton states separate from the background disorder states, even if the disorder states dominate the linear absorption signal *(20-21)*. Here we use single-frequency continuous-wave excitation lasers to perform degenerate ($\omega_{pump} = \omega_{probe}$) coherent pump-probe (CPP) measurements. The degenerate CPP spectrum reveals a series of resonances that are assigned to quantum-confined exciton states.

Finally, we perform non-degenerate CPP measurements using a fixed energy probe beam and variable energy pump beam. *Instead of the expected excitonic spectral hole burning response that is normally observed in ensembles of quantum-confined excitons (22-23), we find that the nonlinear signal follows the featureless PLE spectrum when the pump energy exceeds the probe and then decays rapidly for pump energies less than the probe energy*. Using a model based on the optical Bloch equations, we find that this behavior can be explained by invoking incoherent coupling between optically excited electron-hole pairs in the background disorder states and the exciton states. The nonlinear signal for high pump energies is then provided by rapid non-radiative decay (compared to other population decay rates such as spontaneous emission) from the disorder states into the excitons, resulting in state filling of the excitons and a decrease in probe absorption (saturation). This finding suggests that optical excitation of background disorder can effectively increase the PL brightness of excitons in the system by providing a higher rate of exciton state filling compared to systems with no disorder.

Experiments were performed on an ensemble of site defined InGaN disks in GaN nanowires grown on a GaN template on sapphire substrate using plasma-assisted molecular beam epitaxy (PAMBE). Each nanowire contains ∼ 660 nm GaN, 8 InGaN (3 nm)/GaN (3 nm) vertically aligned quantum disks, and ∼ 5 nm GaN capping layer. The GaN nanowires were grown with a uniform diameter on selective sites on a GaN/sapphire substrate using a patterned titanium film layer as a growth mask as shown in Fig. 1. The nanowires are vertically aligned and highly uniform. The GaN nanowires were grown at a substrate temperature of 830 °C with a gallium (Ga) beam equivalent pressure (BEP) of $2.1 \times 10^{-7}$ Torr and a nitrogen flow rate of 0.55 standard cubic centimeters per minute (sccm). During growth of the InGaN disk/GaN barrier regions, the temperature was reduced to 640 °C with a Ga BEP of $1.5 \times 10^{-8}$ Torr and In BEP of $1.6 \times 10^{-7}$ Torr, respectively. The GaN nanowire is tapered at the end as part of the growth procedure so that several of the InGaN disks within each nanowire have varying diameters. Further details of the growth procedure can be found in Refs. *(12, 25)*. For optical measurements, the sample is maintained at 10 K using a Janis ST-500 cold finger cryostat.

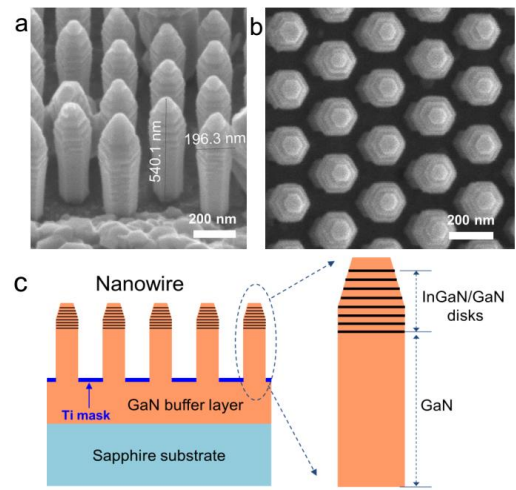

FIGURE 1: (a.) Elevation-view and (b.) plan-view SEM images of the DINW samples. (c.) Cross-sectional view of the sample layers. As part of the growth process, the ends of the DINWs have a tapered design so that the disk diameter varies along the growth direction.

To collect PL data, the sample is excited in a backward geometry by a continuous-wave 3.06 eV diode laser. This laser is replaced by a tunable laser for PLE measurements although in principle an incoherent white light source with a tunable filter can also be used. The excitation lasers are focused onto the sample using an NA = 0.55 microscope objective, resulting in a spot size ~1.5 $\mu m^2$. The areal density of the DINWs is such that ~200-300 DINWs are excited within the optical field of view. The PL spectrum was collected using a nitrogen-cooled CCD camera. Each PLE data point was collected by integrating the total luminescence counts over 1 second within the spectral window ~1.77-1.85 eV using the CCD to collect the spectrum.

To perform CPP measurements, two separately tunable, continuous-wave lasers excite the sample. For degenerate CPP (DCPP) measurements, the energies of the two beams are kept the same. The pump beam and the collinear but non-overlapping probe beam are focused to the same sample spot using the microscope objective used for PL measurements. The differential transmission (dT/T, where T is the sample transmittance) of the probe beam is monitored to detect the changes in sample absorption properties induced by the pump beam using phase-sensitive detection. The reference phase is set so that a positive dT/T signal corresponds to a decrease in probe beam absorption due to the pump beam (saturation). Physically, the saturation signal occurs when electron-hole pairs are excited into an energy state by the pump beam that subsequently reduces the absorption of a probe beam tuned to the same energy state. The lowest order contribution to the differential transmission signal of interest is proportional to the imaginary component of the third order (in the field, $\chi^{(3)}(\omega_{probe} = \omega_{probe} + \omega_{pump} - \omega_{pump})$) nonlinear optical response of the sample that is homodyne detected with the probe beam. We verify that we are in the $\chi^{(3)}$ limit by ensuring that the nonlinear signal strength scales linearly with both the pump and probe beam intensities, and that the measured physical constants do not vary with intensity dependence.

The PL and PLE spectra are shown in Fig. 2. The PL spectrum (black) shows a broad resonance while the PLE spectrum (red) is dominated by a relatively featureless signal that is closely described by an exponential function of energy, as noted earlier. The exponential energy dependence of the PLE spectrum below the bandedge is a common signature of inhomogeneous broadening from disorder states that has been observed frequently in InGaN systems and is similar to an Urbach tail *(26-28)*. It is typical for the linear absorption spectrum in InGaN systems to be dominated by a large density of disorder states rather than excitons, where the density of disorder (and the length of the tail) increases with increasing InN concentration *(28)*.

Although the linear optical signal is typically dominated by disorder, the nonlinear optical signal in InGaN layers is much more sensitive to excitons compared to disorder as noted earlier *(21)*, likely due to a much lower dephasing rate of excitons and longer life time *(21,29)*. Unlike in the mostly featureless PLE spectrum shown in Fig. 2, DCPP data shows a series of resonances that are peaked in amplitude around 2.05 eV and slowly roll off as a function of energy. Based on similar results in previous studies *(20-21)*, we attribute the resonant structures observed in the DCPP spectrum to exciton transitions from different sized DINWs within the ensemble. Energy separations between the exciton states is ~10s of meV and can be accounted for by the variation in DINW diameter due to the tapered design of the DINW *(2,9,12)*. Energy separations of ~10s of meV can also be expected between exciton excited state transitions in the system as well *(9)*. Each of the intensity-independent regions in Fig. 3a therefore likely represents a group of DINWs in the sample with approximately the same diameter that are at the same position along the growth axis of the nanowire or an excited state transition within a group of DINWs. Further support of this interpretation will be provided shortly from comparison of the DCPP data to the intensity-dependent PL spectrum. We rule out that the resonances observed in the nonlinear spectrum are due to etalon effects because the DCPP data shows similar resonances to the modulated absorption (MA) *(30)* spectrum of the sample (not shown), where the pump beam is replaced by a 3.06 eV diode. In this measurement, such an etalon effect would depend strongly on the pump energy since the nonlinear signal (dT/T) is normalized to the probe transmission and no strong pump energy dependence of the resonances has been observed.

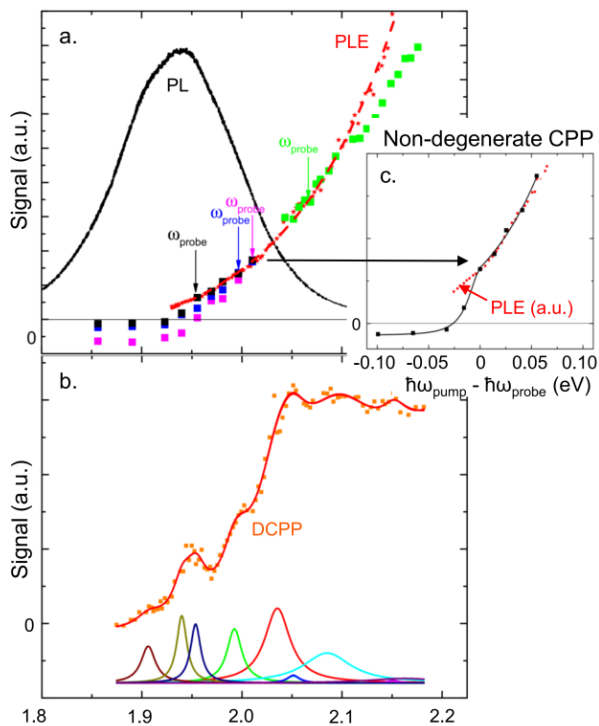

FIGURE 2: a.) PL spectrum (black) plotted with PLE (red), and several non-degenerate CPP spectra with fixed probe energy and scanning pump. The fixed probe energies are shown by the solid arrows. For each fixed probe energy, the corresponding non-degenerate CPP spectrum is color-coded to match the solid arrow. The data are individually normalized to facilitate comparison of the qualitative behavior. The non-degenerate CPP data is scaled so that the $\omega_{pump} = \omega_{probe}$ point coincides with the PLE spectrum. b.) Degenerate CPP ($\omega_{pump} = \omega_{probe}$) fit with a total of 9 exponentially weighted Lorentzians as per Eq. 5. Each Lorentzian represents an exciton transition from a group of DINWs with similar growth properties. c.) Non-degenerate CPP data for a fixed probe energy of 1.95 eV fitted to the indirect term from eq. 5 (black points and line) along with the PLE data (red points) plotted as a function of pump-probe detuning.

The non-degenerate CPP data for fixed probe beam energies and as a function of scanning pump energy is also shown in Fig. 2. We find that the saturation of the probe absorption due to the presence of the pump (positive dT/T) increases rapidly as a function of energy for $\omega_{pump} \geq \omega_{probe}$ and shows a qualitatively similar dependence on energy as the PLE spectrum. For $\omega_{probe} > \omega_{pump}$, the CPP signal tends to decrease more rapidly than the PLE signal and has an additional negative dT/T (an increase in absorption) offset that is observed once the pump beam is sufficiently detuned from the probe. The measurement is repeated for several different fixed probe energies and the same qualitative behavior is observed in each case, however the magnitude of the negative dT/T background signal generally increases as a function of fixed probe energy. A negative dT/T offset has been observed in previous nonlinear spectroscopy studies of self-assembled DINW samples *(21)*. The origin of the negative offset, which can be observed in degenerate CPP spectra as well *(19)*, is still under investigation, however it may be an effect related to bandgap renormalization or screening *(21,31-32)*. A zoomed-in view of the non-degenerate CPP spectrum for a fixed probe near 1.95 eV is shown in Fig. 2c along with the PLE spectrum. The data is fit using a model based on the optical Bloch equations and will be discussed further later on.

In addition to the above nonlinear optical data, the CPP spectrum also features narrow (sub-μeV) resonances at zero pump-probe detuning that are caused by coherent population pulsations (not shown). The nature of the population pulsation resonances has been explored in a previous work featuring an identical sample at room temperature *(33)*. The width of the population pulsation resonance gives the population decay rate of the state being probed. It is interesting to note that the width of population pulsation resonances that gives a population decay time ~5-10 ns in the sample is virtually independent of temperature from 10 K-300 K, which likely reflects the large exciton binding energy in this system compared to materials such as III-As.

To determine whether the PL emission in our system originates from excitons or disorder states, we consider the excitation intensity dependence of the PL emission energy. It has been shown that emission from exciton states in individual, localized DINW systems show no energy shift as a function of excitation intensity and effects such as charge screening of the internal electric fields that would result in carrier density-dependent shifts are insignificant *(2,34)*. On the other hand, the intensity-dependent PL of localized disorder states in planar InGaN quantum wells normally shows a progressive blue shift as a function of increasing excitation intensity that is attributed to a combination of intensity-dependent disorder state filling and internal electric field screening *(34)*.

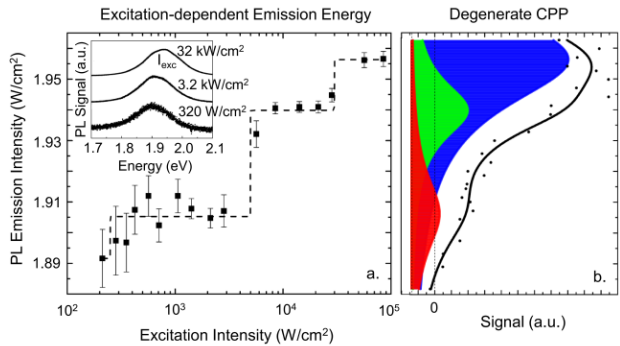

FIGURE 3: (a.) PL peak energy plotted as a function of excitation intensity. The error bars represent variation in the measurement possibly caused by small sample vibrations. The inset shows the PL spectrum for 3 different orders of magnitude in excitation intensity. The dashed line is a fit to the simple model. (b.) DCPP spectrum with a 3-Lorentzian fit in the same energy window as 3a. The peak energy of the Lorentzians are the same energy as the excitation intensity-independent regions in 3a.

The excitation intensity dependence of the PL emission energy from the ensemble DINW sample shown in Fig. 3a exhibits discrete blue shifts in the sample emission energy followed by independence of sample emission energy as a function of increasing excitation intensity. The excitation intensity-independent parts of the spectrum in Fig. 3 are consistent with emission from quantum confined exciton states as noted earlier *(2)*. The discrete blue shifting behavior in Fig. 3a can be explained by competition from different groups of excitons where lower energy exciton states have a higher relative brightness at lower excitation intensities but a lower saturation intensity and maximum brightness compared to higher energy exciton states. Similar behavior has been observed in a previous study *(21)* where larger diameter DINWs (with a higher emission energy) have a higher electroluminescence brightness for high injection currents. This is contrary to simple two-level dynamics but can be accounted for in the model discussed in the Supplemental Material. For a guide to the eye, the data in Fig. 3a is fit to a series of step functions, where the vertical transition to each step represents excitation intensities for which the PL brightness of two different groups of disks is the same.

In Fig. 3b we compare the DCPP data to the intensity-dependent PL data from Fig. 3a within the same energy window. As a simple approximation, we find that the DCPP data can be fit to a series of Lorentzians, each representing nonlinear absorption from exciton transitions. Later on we will modify this model to account for the effects of coupling between disorder states and excitons that accounts for the relative amplitudes of the resonances. We find that the center energy of the Lorentzians from the fit in Fig. 3b shows good agreement with the excitation intensity-independent energies from Fig. 3a. This data shows a consistency in our interpretation of the nonlinear absorption signal being mainly sensitive to exciton states rather than absorption from the disorder states.

To understand the nondegenerate nonlinear data, we consider a model of the system in which electron-hole pairs can be excited into background disorder states and subsequently decay non-radiatively into lower energy exciton states as shown in Fig. 4a. Figure 4b is the model for the usual case with no significant coupling between the disorder states and the exciton. The non-radiative decay process could occur from the emission of acoustic phonons, for example *(35-36)*. In our model, we assume that the system is excited with pump and probe beams where the probe beam is also resonant with a homogeneously broadened exciton state |X> and that the pump beam is allowed to vary in energy. We will calculate the third order nonlinear absorption spectrum of the system perturbatively using the optical Bloch equations in the density matrix picture *(33, 37)*.

In the density matrix picture, the equations of motion for the density matrix terms of interest are given in an interaction representation in the dipole approximation by *(37)*

$$\dot{\rho}_{XG}{}^I = i\chi_\mu(\rho_{XX} - \rho_{GG}) - \gamma\rho_{XG}{}^I \quad (1)$$

$$\dot{\rho}_{XX} = i\chi_\mu{}^*\rho_{GX}{}^I + c.c. - \Gamma_X\rho_{XX} + \Gamma_D\rho_{DD} \quad (2)$$

where $\chi_\mu = -(\boldsymbol{\mu}_{XG} \cdot \boldsymbol{\varepsilon}_\mu)E_\mu/2\hbar$. Here $E_\mu$ represents the applied field with polarization vector $\boldsymbol{\varepsilon}_\mu$ and $\boldsymbol{\mu}_{XG}$ is the transition dipole matrix element.

The decay rates are given by $\Gamma_X$, the total population decay rate of the disorder states $\Gamma_D$, and $\gamma = \Gamma_X/2 + \Gamma_{dec}$, where $\Gamma_{dec}$ is the rate of decay of exciton coherence due to pure dephasing (i.e., physical processes that do not result in loss of population). Pure dephasing can occur in semiconductor systems due to effects such as exciton-phonon interactions or fast charge fluctuations *(33, 38-39)*. The equation of motion for the disorder state population is given in the rate equation approximation as

$$\dot{\rho}_{DD} = -W(\rho_{DD} - \rho_{GG}) - \Gamma_D \rho_{DD} \quad (3)$$

where W is the optical pumping rate from the ground state into the disorder state continuum. As an approximation, to focus on the behavior and quantitative details, we assume the decay rate from the disorder states to exciton states is much greater than recombination from the disorder states to the ground state and hence we do not include decay from the disorder states to the ground state. The measurement is proportional to the imaginary part of the third order off-diagonal density matrix element $\rho_{XG}^{(3)}(\omega_{pump}, \omega_{probe})$. From perturbation theory, the third order off-diagonal term (in the applied optical fields in the direction of probe field) is given by *(37)*

$$\rho_{XG}^{(3)}(\omega_{probe}, \omega_{pump}) = \frac{i\chi_{probe}}{\gamma + i(\omega_X - \omega_{probe})}\left(\rho_{XX}^{(2)}(\omega_{pump}) - \rho_{GG}^{(2)}(\omega_{pump})\right) \quad (4)$$

where $\rho_{XX}^{(2)}$ and $\rho_{GG}^{(2)}$ are terms that are second order in the applied fields corresponding to the populations of the exciton and ground state.

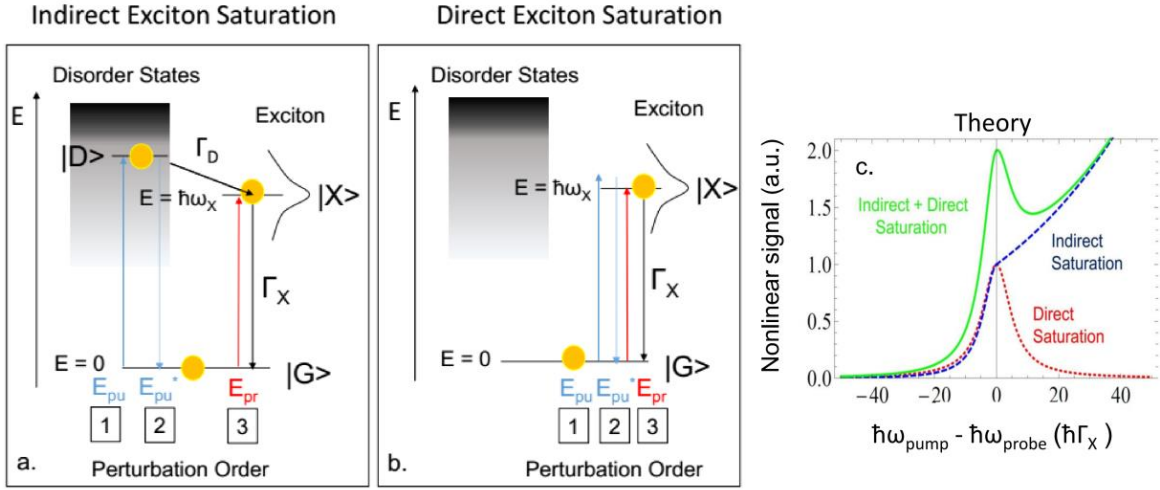

FIGURE 4: (a.) Here we show a model for the energy levels in the system along with the perturbation sequence for indirect exciton saturation. The exciton state is represented by |X> with energy $\hbar\omega_x$ while the disorder states that are resonant with the pump beam are represented by |D>. The yellow circles represent states that have a non-zero population of electron-hole pairs in second order. (b.) Corresponding energy level diagram and perturbation sequence for direct exciton saturation. (c.) Theoretical nonlinear response from the imaginary part of Eq. 5.. The indirect term (first term on the right-side of Eq 5 and corresponding terms in Eqs. 6-7 (green dashed) is shown separate from the direct term (second on the right-hand side of Eq. 5, red dots).

We consider 2 different excitation pathways that contribute to the third order signal via the second order population terms in Eq. 4: the first represents excitation of electron-hole pairs in the background disorder states in first and second order of perturbation theory that can non-radiatively decay into the exciton state and reduce the absorption of the exciton state in third order (indirect exciton saturation) as shown in Fig. 4a. The second non-linear term represents direct nonlinear absorption of the excitons in which the pump and probe beam excites an exciton population in first and second order that directly saturates the absorption of the probe beam in third order as shown in Fig. 4b. Here we will not consider the population pulsation terms when the probe and pump beams act in first and second order, respectively as these were the subject of a previous study *(33)* as noted earlier. Coherent coupling between the excitons and background disorder states or direct excitation of background disorder states are likely to be small compared to the excitonic nonlinear absorption terms and will also be ignored. The resulting third order off-diagonal density matrix element from the two pathways is given by

$$\rho_{XG}^{(3)}(\omega_{probe}, \omega_{pump}) = \chi_{probe}\left(\rho_{DD}^{(2)}(\omega_{pump})\frac{\Gamma_D(\omega_{pump})}{\Gamma_X}\right)\frac{i}{\gamma+i(\omega_X-\omega_{probe})} + \frac{\chi_{probe}|\chi_{pump}|^2 \times 2i}{[\gamma+i(\omega_X-\omega_{probe})][\gamma^2+(\omega_X-\omega_{pump})^2]} \quad (5)$$

where the first term on the right-hand side of Eq. 5 represents the indirect exciton saturation and the second term represents the direct exciton saturation. Based on our interpretation of the PLE spectrum in terms of Urbach tail physics, we take the second-order disorder state population from the indirect term $\rho_{DD}^{(2)}(\omega_{pump})$ to be given by the exponential function

$$\rho_{DD}^{(2)}(\omega_{pump}) = N_0 e^{\frac{\omega_{pump}}{\omega_0}} \quad (6)$$

The decay rate from the disorder states clearly depends on the wavelength. For example, for a pump wavelength much longer than the central resonant wavelength of exciton, the decay rate is 0 (in the absence of processes such as phonon assisted absorption) because of energy conservation. Therefore, we assume a phenomenological pump-wavelength dependent decay rate from the background disorder states to the exciton to be:

$$\Gamma_D(\omega_{pump}) = \Gamma_{D0}\left(\frac{\gamma}{2}\right)^2 \frac{1}{(\omega_{pump}-\omega_X)^2+(\gamma/2)^2}\left(1-\Theta(\omega_{pump}-\omega_X)\right) + \Gamma_{D0}\Theta(\omega_{pump}-\omega_X) \quad (7)$$

where the decay is weighted by a normalized Lorentzian that describes the linear absorption of the exciton state for $\omega_{pump} < \omega_{probe}$ due to the homogeneously broadened exciton transition. Θ is the usual Heaviside function. Because the measurement is taken at low temperature, we do not consider processes such as phonon absorption that can provide an addition source of exciton saturation when $\omega_{pump} < \omega_{probe}$ to be significant.

In Fig. 4c, we plot the theoretical direct and indirect saturation components (red and blue lines) as well as the total signal expected from the imaginary part of Eq. 5 (green line) assuming that $\Gamma_{dec} = 10\Gamma_X$. At zero pump-probe detuning (both pump and probe are resonant with the exciton state), the direct term shows a resonance while the indirect term shows a monotonically increasing signal as a function of pump energy that follows the same dependence as the exponential PLE signal for $\omega_{pump} > \omega_{probe}$. If both the direct and indirect saturation terms contribute equally, the signal shows a peak at zero pump-probe detuning that is skewed by the exponential density of states term.

The non-degenerate CPP data in Fig. 2b shows no evidence of resonant structures that are characteristic of the direct exciton excitation response from Fig. 4c. Instead, the data indicates that the nonlinear response is mostly dominated by the indirect saturation term that results in a monotonic increase in signal as a function of pump energy with respect to the fixed probe energy. Indeed, the non-degenerate CPP signal for ħω_probe = 1.95 eV, nearly on resonance with an exciton transition from the DCPP data shown in Fig. 2b, shows a good fit to the indirect exciton saturation response as shown in Fig. 2c. From the fit, we estimate the homogeneous linewidth of the exciton state to be ~15 meV, consistent with previous reports at room temperature *(33)*. For this system, the fit therefore provides similar information to spectral hole burning *(37)* of the inhomogeneously broadened exciton ensemble. The dominance of the indirect saturation term over the direct term indicates that the "in" term $\Gamma_D\rho_{DD}^{(2)}$ largely exceeds both the direct optical pumping rate into the exciton state from the pump beam and the total decay rate of the exciton $\Gamma_X$, hence the non-radiative decay is rapid compared to these terms. Note that the signal for $\omega_{pump} < \omega_{probe}$ could contain additional peaks shifted from zero pump-probe detuning if the probe energy was fardetuned from the exciton peak or if multiple coupled exciton states were probed, however we expect that the additional peaks in this case will be relatively small if the indirect term is dominant over the direct term.

To further show consistency between the data and the above model, we fit the DCPP signal from Fig. 2b using the indirect term from Eq. 5 with $\omega_{pump} = \omega_{probe}$, assuming that the indirect term is dominant over the direct term from Eq. 5. In this case, the DCPP signal is given by a series of exponentially weighted Lorentzians as a function of energy, where each Lorentzian represents a group of excitons from DINWs with the same morphological properties or groups of exciton excited states. In Fig. 2b, we fit the DCPP data to a series of 9 exponentially weighted Lorentzians. We find that the value for $1/\omega_0$ from Eq. 5 extracted from the fit ($0.076 \pm 0.005$) shows some discrepancy from a single exponential fit to the PLE data shown in Fig. 2a, where $1/\omega_0 = 0.109 \pm 0.003$. This discrepancy is possibly related to an incomplete knowledge of the exciton energy distribution in the system, resulting in a fit in Fig. 2b with too few resonances. It is also possible that the fit can be influenced by higher energy transitions outside of the laser tuning range. Future experimental and theoretical work should focus on understanding exciton energy levels on the single DINW level. Shown for reference in Fig. 2b is the nonlinear absorption from the exciton transitions in the absence of disorder state coupling (i.e. the direct excitation term). The three lowest energy Lorentzians can be compared directly to the Lorentzians from the fit in Fig. 3b. It is interesting to note that the higher energy transitions are generally broader compared to the lower energy transitions, which could reflect a faster nonradiative decay from higher energy states (i.e. excited states).

We also consider population transfer from exciton excited states to lower energy exciton states such as by Forster *(40)* or Dexter *(41)* energy transfer that can add to the dT/T signal. This would likely lead to additional resonances in the nondegenerate CPP response at higher pump energies compared to the probe that reflect the density of exciton excited states that may be visible in measurements such as DCPP. If this type of population transfer is dominant, we would expect the non-degenerate signal to show a similar energy dependent roll-off as the DCPP for high pump energies. In Fig. 2b, the non-degenerate spectrum is probed at a fixed probe energy exceeding the peak of the DCPP spectrum in Fig. 2b (green data points). The non-degenerate signal is again qualitatively similar to the PLE spectrum for $\omega_{pump} \geq \omega_{probe}$ up to approximately $\omega_{pump} = 2.115$ eV, where the dT/T signal shows a small peak with a corresponding small peak in the DCPP signal. The presence of a peak in signal suggests that population excited into exciton excited states can contribute slightly to the CPP signal, however for $\omega_{pump} > 2.115$ eV we find that the signal shows a monotonic increase as a function of energy, whereas the DCPP signal levels off and begins to slowly decrease with increasing pump energy. This indicates that the pump-probe signal is mostly determined by population decay from the background disorder states into the excitons rather than exciton excited states.

In summary, the above results indicate that optical excitation of the sample results in a steady-state lowest order (second order in the applied field) exciton population that is primarily populated by via non-radiative coupling from higher energy disorder states. As discussed in the Supplemental Material, we expect the PL brightness to scale as $N\Gamma_{XR}$ for high excitation intensities, where N is proportional to the total population of disorder states that are coupled to the exciton and $\Gamma_{XR}$ is the radiative decay rate of the exciton. This shows that the emission efficiency of InGaN DINWs at high excitation intensities might be improved by the incorporation of disorder in the system, a result that was similarly suggested in the context of quantum dot-deep level trap interactions *(42)*. As noted in the Supplemental Material, it is possible that larger diameter DINWs (with lower emission energies in this sample) have a higher value of N (while possibly a higher non-radiative decay rate) compared to lower diameter DINWs due to a larger surface area. This can be used to explain the intensity-dependent PL behavior observed in Fig. 3a. It should be noted that fast charge transfer processes can increase the linewidth of the excitons by increasing the exciton decoherence *(33)*, which can be problematic for certain applications such as quantum information processing.

## AUTHOR INFORMATION

**Corresponding Author**

* crnelson@umich.edu

**Present Addresses**

§ Currently at Intel Corporation.

## Author Contributions

The manuscript was written through contributions of all authors. / All authors have given approval to the final version of the manuscript.

## Funding Sources

This work was funded by the National Science Foundation through CPHOM (Grant No. DMR 1120923) NSF 170862 and through US Army Research Office (W911NF-17-1-0109).

# Supplemental Material

**Intensity-Dependent Photoluminescence (PL) Spectrum**

The peak energy of the PL emission in the sample shows discrete blue shifts as a function of excitation intensity, as shown in Fig. 3 of the main text. To understand this effect, we consider emission from 3 different disks, each with different exciton transition energies as shown in Fig. S1. As noted in the main text, the smaller diameter disks are expected to have lower exciton emission energies compared to larger diameter disks, therefore we take the diameters of the disks in Fig. S1 to be decreasing from left to right. Here we will use the optical rate equations to calculate the steady-state PL emission as a function of excitation intensity from a 3.06 eV pump beam that is tuned into higher disorder/continuum states of each disk (represented by in Fig. S1 by the ket $|D_i\rangle$).

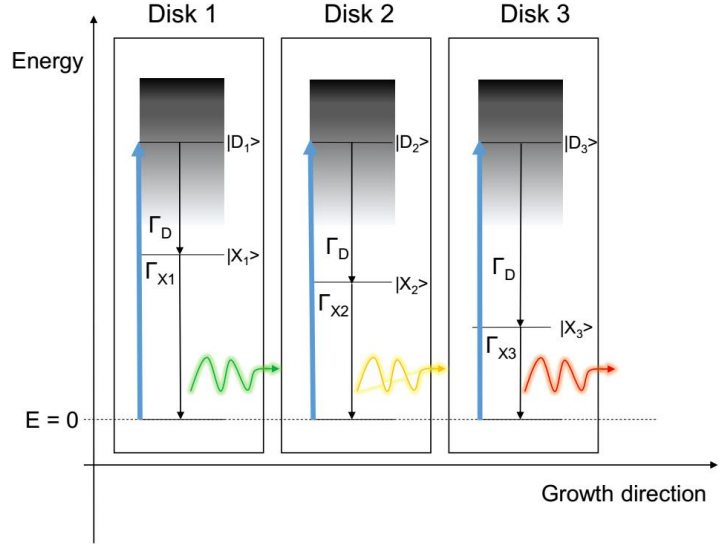

Figure S1: Energy level diagrams of 3 different DINWs, where it is assumed that the lower diameter DINWs have a lower emission energy compared to larger diameter DINWs. Electron-hole pairs are optically excited by a pump beam into the continuum energy levels and subsequently decay non-radiatively into the exciton states before radiatively recombining.

The steady-state optical rate equations in the above system for each disk are given by

$$\dot{D}_i = W(L_i - U_i) - \Gamma_D U_i = 0$$
$$\dot{L}_i = -W(L_i - U_i) + \Gamma_{Xi} X_i = 0$$
$$\dot{X}_i = \Gamma_D U_i - X_i \Gamma_{Xi} = 0$$
$$X_i + D_i + L_i = N_i \quad (i = 1,2,3)$$

(S1)

where W is the optical pumping rate (proportional to the excitation intensity $I_{exc}$) and $D_i$, $L_i$ and $X_i$ are the populations of the disorder states, ground state and exciton state, respectively, in each of the DINWs and $N_i$ represents the total population of disorder states that are coupled to the exciton as well as the population of the exciton state. In this model, we ignore recombination from the disorder states to the crystal ground state and we assume that $\Gamma_D \gg \Gamma_i$ based on the non-degenerate nonlinear and PLE data from the main text. We further assume that $\Gamma_D$ is approximately the same for each DINW. Under these assumptions, the steady-state solution to equation S1 is approximately given by

$$X_i \approx \frac{N_i W}{W + \Gamma_{Xi}}$$
(S2)

The PL brightness $B$ is given by $B = \Gamma_{XiR} X_i$, where $\Gamma_{XiR}$ is the radiative decay rate of the exciton and $\Gamma_{Xi} = \Gamma_{XiR} + \Gamma_{XiNR}$, where $\Gamma_{XiNR}$ is the non-radiative decay rate of the exciton. In the limit of low excitation intensity, the PL brightness is approximately given by

$$B_{low\ intensity} \approx \frac{N_i W \Gamma_{XiR}}{\Gamma_{Xi}} = \frac{N_i W \Gamma_{XiR}}{\Gamma_{XiR} + \Gamma_{XiNR}} \sim \frac{N_i I_{exc} \Gamma_{XiR}}{\Gamma_{XiR} + \Gamma_{XiNR}}$$
(S3)

where $I_{exc}$ is the excitation intensity of the 3.06 eV beam, while at high intensity the brightness is approximately given by

$$B_{high\ intensity} \approx N_i \Gamma_{XiR}$$
(S4)

For different disks, the values of $N_i$, $\Gamma_{XiR}$ and $\Gamma_{XiNR}$ can be varied to give the intensity-dependent energy shifting behavior observed in Fig. 3 of the main text. Particularly, disks with a significantly higher value of $\Gamma_{XiNR}$ and a higher value of $N_i$ or $\Gamma_{XiR}$ compared to other disks can have a lower brightness at low excitation intensities but a much higher saturation intensity that results in a higher brightness at high excitation intensities compared to disks with lower values of $\Gamma_{XiNR}$ and $N_i$ or $\Gamma_{XiR}$ from equations S3 and S4.

For smaller diameter DINWs, diffusion of InN atoms from the lateral sidewalls of the DINW structure toward the center of the DINW leads to a higher density of InN in the center of the disk compared to larger diameter DINWs *(S1)*. As a result of the higher InN concentration, excitons in the center of smaller diameter DINWs are expected to have a higher quantum confinement energy (and lower emission energy) that can lead to a lower non-radiative recombination rate compared to larger diameter DINWs. In larger diameter DINWs, the larger surface area of the disk could result in a higher total number of total disorder states within the DINW as a larger number of InN atoms are impinging on the DINW surface during the growth procedure. It is possible that this could result in a higher number of disorder states that are coupled to the exciton N for larger diameter DINWs. Currently the relative radiative decay rates between excitons in different sized DINWs in selective area samples is unknown. It is possible however that lower diameter DINWs with a smaller value of $\Gamma_{XNR}$ and N may have a higher PL brightness at low excitation intensities and lower brightness at high excitation intensities compared to larger diameter DINWs based on the above discussion.

(S1): Sekiguchi, H.; Kishino, K.; Kikuchi, A. *Appl. Phys. Lett.* **2010**, *96*, 231104.